\documentclass[aps,pre,showpacs,onecolumn]{revtex4}
\usepackage{bm}
\usepackage{graphicx}
\bibstyle{approve.bib}
\usepackage{amssymb}
\usepackage{epstopdf}
\usepackage{color}
\usepackage{gensymb}

\newcommand{\be}{\begin{equation}}
\newcommand{\ee}{\end{equation}}
\newcommand{\bea}{\begin{eqnarray}}
\newcommand{\eea}{\end{eqnarray}}
\newcommand{\lan}{\left\langle}
\newcommand{\ran}{\right\rangle}
\newcommand{\br}{\mathbf{r}}
\newcommand{\ba}{\mathbf{a}}
\newcommand{\bE}{\mathbf{E}}

\newcommand{\bp}{\mathbf{p}}

\newcommand{\bq}{\mathbf{q}}
\newcommand{\bP}{\mathbf{P}}

\newcommand{\bom}{\mathbf{\Omega}}

\newcommand{\bv}{\mathbf{u}}
\newcommand{\e}{\varepsilon}

\newcommand{\tv}{\tilde{v}}

\newcommand{\tphi}{\tilde{\phi}}

\newcommand{\baq}{\bar{q}}

\begin{document}

\title{Dielectric anisotropy in polar solvents under external fields}

\author{Sahin Buyukdagli$^{1}$\footnote{email:~\texttt{Buyukdagli@fen.bilkent.edu.tr}}}
\affiliation{$^{1}$Department of Physics, Bilkent University, Ankara 06800, Turkey}
\date{\today}

\begin{abstract}
We investigate dielectric saturation and increment in polar liquids under external fields. We couple a previously introduced dipolar solvent model to a uniform electric field and derive the electrostatic kernel of interacting dipoles. This procedure allows an unambiguous definition of the liquid dielectric permittivity embodying non-linear dielectric response and correlation effects. We find that the presence of the external field results in a dielectric anisotropy characterized by a two-component dielectric permittivity tensor. The increase of the electric field amplifies the permittivity component parallel to the field direction, i.e. dielectric increment is observed along the field. However, the perpendicular component is lowered below the physiological permittivity $\e_w\approx77$, indicating dielectric saturation perpendicular to the field. By comparison with Molecular Dynamics simulations from the literature, we show that the mean-field level dielectric response theory underestimates dielectric saturation. The inclusion of dipolar correlations at the weak-coupling level intensify the mean-field level dielectric saturation and improves the agreement with simulation data at weak electric fields.  The correlation-corrected theory predicts as well the presence of a metastable configuration corresponding to the antiparallel alignment of dipoles with the field. This prediction can be verified by solvent-explicit simulations where solvent molecules are expected to be trapped transiently in this metastable state.
\end{abstract}

\date{\today}
\maketitle

\section{Introduction}

The importance of water as a regulator of biological processes has been the driving force behind the intense research on the electrostatic properties of this solvent. In particular, the non-linear dielectric response of water has been exhaustively scrutinized for the past nine decades. The seminal idea from Debye~\cite{Debye1929} on the saturation of the water polarizability by external fields was reconsidered by various theories. To name the milestones, one can mention Onsager's dielectric cavity concept modelling the central dipole as a cavity in a continuum dielectric fluid~\cite{Onsager1936}, the statistical mechanical approach from Kirkwood relaxing this continuum approximation~\cite{Kirkwood1939}, and Booth's celebrated law extending the previously mentioned models to the non-linear dielectric response regime of strong fields~\cite{Booth1951}. 

Booth's law states that the solvent dielectric permittivity $\e$ drops with the externally applied field $E$ according to the non-linear relation 
\be\label{nl}
\e=\eta^2+c_1\frac{\mathrm{L}(c_2E)}{E}, 
\ee
where $\eta$ stands for the optical retractive index of air, $c_{1,2}$ are model parameters characteristic of water, and $L(x)=\mathrm{coth}(x)-1/x$ is the Langevin function. Eq.~(\ref{nl}) predicting the reduced polarizability of solvent molecules in strong external fields has been applied to various soft matter problems, such as the swelling pressure in clays~\cite{Curry1992}, the interaction of thin films immersed in electrolytes~\cite{Basu1994}, ionic hydration~\cite{Sandberg2002}, the electrostatic coupling between DNA strands~\cite{Gavryushov2007}, and the electrostatics of protein channels~\cite{Arzo2008}. Booth's formalism was also verified by Molecular Dynamics (MD) simulations of dipoles in strong fields~\cite{Yeh1999,Fulton2009}. Improved versions of the theory based on the Ornstein-Zernike equation were developed as well in Refs.~\cite{Wertheim,Hoye,Stell,Szalai2009}. 

The standard relation between the polarization field $P$ and the dielectric permittivity 
\be\label{bo}
\e=1+\frac{4\pi P}{E}
\ee
used in Booth's theory~\cite{Booth1951} and in MD simulations~\cite{Yeh1999} presents some ambiguity. Strictly speaking, the definition of the dielectric susceptibility as the ratio of the polarization field and the external field is valid exclusively in the linear response regime. Furthermore, such an algebraic equation neglects non-local dielectric response effects stemming from the extended charge structure of water molecules. This point was considered in Ref.~\cite{NLPB1} where we had investigated the relation between non-locality and solvent charge structure at the mean-field (MF) level. Most importantly, the definition of the dielectric permittivity~(\ref{nl})  as a scalar quantity oversimplifies the underlying physics. Indeed, the presence of the external field is expected to break the spherical symmetry in the polar liquid. This should translate in turn into a tensor form of the dielectric permittivity that cannot be expressed as a scalar function.

Motivated by these points, we reconsider in this work the dielectric saturation problem within a microscopic polar liquid model recently introduced in Ref.~\cite{NLPB1}. This non-local model is based on a field-theoretic formulation of a polar liquid composed of finite-size dipoles.  Similar scalar field theories have been previously applied to charged fluids composed of point-ions~\cite{Podgornik1988,Netz2000,Lau2008}, point-dipoles~\cite{dunyuk,orland1}, and charges with internal structure~\cite{bohdip,Lue2}. The solvent-explicit electrolyte model of Ref.~\cite{NLPB1} was also extended beyond MF level in order to investigate the hydration of polarizable ions in polar liquids~\cite{NLPB2} and solvent response in nanoslits~\cite{Buyuk2014II}. We develop in Section~\ref{theory} the one-loop theory of interacting dipoles subject to an external field. This provides us with the electrostatic kernel embedding the non-local dielectric response properties of the liquid beyond mean-field level. We have previously applied similar one-loop approaches for dielectric continuum liquids to ion partition in nanochannels~\cite{Buyuk2012,Buyuk2013} and polymer translocation through nanopores~\cite{Buyuk2014III,Buyuk2015} with extensive comparisons with Monte-Carlo simulations and experimental data. The inclusion of non-locality and correlation effects neglected by the previous formalisms of Ref.~\cite{dunyuk,orland1,bohdip,Lue2} is the key advance of the present approach. The infrared limit of the electrostatic kernel yields the two-component dielectric permittivity tensor of the liquid. In terms of this  permittivity tensor, we characterize in Section~\ref{res} the dielectric increment and saturation effects induced by the electric field as well as the role played by dipolar correlations on the corresponding dielectric anisotropy. The limitations of our model are discussed in the Conclusion.

\section{Weak-coupling theory of interacting dipoles}
\label{theory}

\begin{figure}
\includegraphics[width=.6\linewidth]{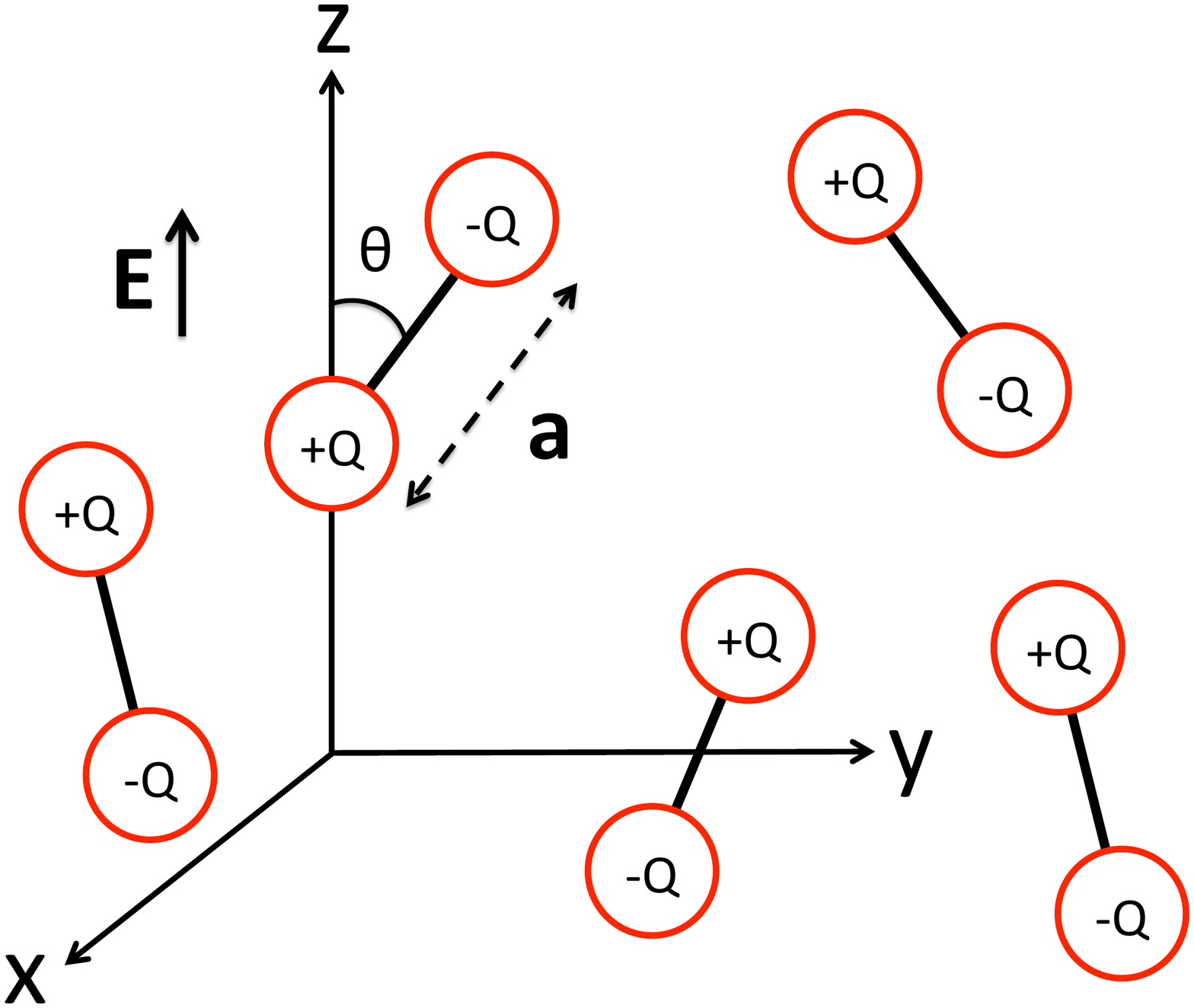}
\caption{(Color online) Composition of the solvent molecules filling the bulk medium where an external electric field $\bE$ is applied along the $z-$axis. In the present work, the valency of the dipolar molecules is $Q=1$, the solvent molecular size $a=1.0$ {\AA}, and the solvent density $\rho_s=55.0$ M. According to the Debye-Langevin relation~(\ref{dibulk}), at vanishing electric field $\bE=0$, these parameters correspond to the water dielectric permittivity $\e_w\approx77$.}
\label{Fig0}
\end{figure}

In this section, we derive the weak-coupling electrostatic propagator of the polar liquid model introduced in Ref.~\cite{NLPB1} in order to investigate its dielectric response properties under  an externally applied electric field. The geometry of solvent molecules and the composition of the dipolar liquid are depicted in Fig.~\ref{Fig0}. The liquid is composed of electrostatically interacting dipoles with finite size $a$ and elementary charges $\pm Q$ at the edges. In a bulk medium where an external field $\bE(\br)$ is present, the grand-canonical partition function of the solvent model derived in Ref.~\cite{NLPB1} takes the form
\be\label{zg1}
Z_G=\sum_{N_s\geq0}\frac{\Lambda_s^{N_s}}{N_s!}\int \prod_{n=1}^{N_s}\mathrm{d}\br_n\frac{\mathrm{d}\bom_n}{4\pi}e^{-H_c+\int\mathrm{d}\br\;\bE(\br)\cdot\bP(\br)},
\ee
where $\Lambda_s$ is the dipolar fugacity, $N_s$ the number of solvent molecules, and the vectors $\br_n$ and $\bom_n$ are respectively the position vector of the $n^{\rm{th}}$ solvent molecule and the solid angle characterizing its orientation. Furthermore, the electrostatic interaction energy is given by 
\be\label{elpw}
H_c=\int\frac{\mathrm{d}\br\mathrm{d}\br'}{2}\rho_c(\br)v_c(\br,\br')\rho_c(\br')
\ee
with the dipolar charge density
\be
\rho_c(\br)=Q\sum_{n=1}^{N_s}\left[\delta(\br-\br_n)-\delta(\br-\br_n-\ba_n)\right],
\ee
the Coulomb potential $v_c(\br)=\ell_B/r$, and $\ell_B\approx 55$ nm the Bjerrum length in vacuum. We also introduced in Eq.~(\ref{zg1}) the dipole-field coupling in terms of the external field $\bE(\br)$ and the polarization density $\bP(\br)=\sum_{n=1}^{N_s}\bp_n\delta(\br-\br_n)$, where the dipole vector $\bp_n=-Q\ba_n$ points the positive charge of the $n^{\rm{th}}$ solvent molecule.

After performing a standard Hubbard-Stratonovich transformation in order to linearize the quadratic coupling in Eq.~(\ref{elpw}), the grand-canonical partition function takes the form of a functional integral over the fluctuating electrostatic potential, $Z_G=\int \mathcal{D}\phi\;e^{-H[\phi]}$, with the Hamiltonian functional
\be\label{HamFunc}
H[\phi]=\int\mathrm{d}\br\;\frac{\left[\nabla\phi(\br)\right]^2}{8\pi\ell_B}-\Lambda_s\int\frac{\mathrm{d}\br\mathrm{d}\bom}{4\pi}e^{-Q\bE\cdot\ba}e^{iQ\left[\phi(\br)-\phi(\br+\ba)\right]}.
\ee
Because the Hamiltonian functional is non-linear, we cannot integrate the partition function analytically. To progress further, we will compute the partition function and the field theoretic averages perturbatively. This perturbative approach is similar to the one-loop expansion of the grand potential of inhomogeneous electrolyte solutions~\cite{Podgornik1988,Netz2000,Lau2008}. First, we express the Hamiltonian as the sum of its quadratic part in the fluctuating potential $\phi(\br)$ and its non-linear part,
\be\label{exp}
H[\phi]=H_0[\phi]+\delta H[\phi],
\ee
where the quadratic part corresponding to the second order Taylor expansion of Eq.~(\ref{HamFunc}) in $\phi(\br)$ reads
\bea\label{qu1I}
H_0[\phi]&=&\int\mathrm{d}\br\;\frac{\left[\nabla\phi(\br)\right]^2}{8\pi\ell_B}-\Lambda_s\int\frac{\mathrm{d}\br\mathrm{d}\bom}{4\pi}e^{-Q\bE\cdot\ba}\left\{iQ\left[\phi(\br)-\phi(\br+\ba)\right]-\frac{Q^2}{2}\left[\phi(\br)-\phi(\br+\ba)\right]^2\right\}.\nonumber
\eea
In the second integral of Eq.~(\ref{qu1I}), we dropped an irrelevant constant that results from the expansion of the exponential function. From now on, we will assume that the external field is uniform and oriented along the $z$-axis, i.e. $\bE(\br)=E_z\bv_z$. To exploit the corresponding translational symmetry within the $(x,y)$ plane and along the $z$-axis, we expand the potential in Fourier basis as 
\be\label{fr1}
\phi(\br)=\phi\left(\br_\perp, z\right)=\int\frac{\mathrm{d}^2\bq_\perp}{(2\pi)^2}\frac{\mathrm{d}q_z}{2\pi}e^{i\bq_\perp\cdot\br_\perp}e^{iq_zz}\tphi\left(q_\perp,q_z\right),
\ee
where $\bq_\perp$ and $q_z$ are respectively the wave vectors in the reciprocal space of the $(x,y)$ plane perpendicular to the field and the $z$-axis parallel to the field. By inserting this Fourier expansion into Eq.~(\ref{qu1I}), the term linear in the fluctuating potential vanishes and the quadratic Hamiltonian takes the form
\be
H_0[\phi]=\frac{1}{2}\int\frac{\mathrm{d}q_\perp q_\perp}{2\pi}\frac{\mathrm{d}q_z}{2\pi}\tv_0^{-1}(q_\perp,q_z)\;\tphi\left(q_\perp,q_z\right)\tphi\left(-q_\perp,-q_z\right),
\ee
which can be expressed in terms of the potentials in real space as
\be\label{qu1}
H_0[\phi]=\frac{1}{2}\int\mathrm{d}\br\mathrm{d}\br'\phi(\br)v_0^{-1}(\br,\br')\phi(\br').
\ee
In Eq.~(\ref{qu1}), the Fourier transform of the electrostatic kernel $v_0^{-1}(\br,\br')$ reads 
\be\label{fr2}
\tv_0^{-1}(q_\perp,q_z)=\frac{q^2}{4\pi\ell_B}+2Q^2\Lambda_s\frac{\sinh (u)}{u}F(q_\perp,q_z),
\ee
with the wave vector $q^2=q_\perp^2+q_z^2$ and the adimensional external field
\be
u=QaE_z.
\ee
In Eq.~(\ref{fr2}), we also introduced the auxiliary function corresponding to the structure factor of solvent molecules coupled to the external field,
\be\label{f1}
F(q_\perp,q_z)=1-\lan\rm{J}_0(q_\perp a_\perp)\cos(q_za_z)\ran_\theta,
\ee
where we defined the average over dipolar rotations of a general function $T(\theta)$ in the form
 \be
 \lan T(\theta)\ran_\theta=\frac{u}{2\sinh (u)}\int_0^\pi\mathrm{d}\theta\sin\theta\;e^{-u\cos\theta} T(\theta).
 \ee
In Eq.~(\ref{f1}), $\theta$ stands for the polar angle between the electric field and the dipole (see Fig.~\ref{Fig0}), $\rm{J}_0(x)$ is the Bessel  function of the first kind, and the components of the dipolar orientation perpendicular and parallel to the electric field are respectively defined as
\bea
&&a_\perp=a\sin\theta\\
&&a_z=a\cos\theta.
\eea

We will compute the dielectric permittivities from the large distance behaviour of the two point correlation function $v(\br,\br')\equiv\lan\phi(\br)\phi(\br')\ran$. The latter is given by 
\be\label{corr1}
v(\br,\br')=\frac{1}{Z_G}\int \mathcal{D}\phi\;e^{-H[\phi]}\phi(\br)\phi(\br').
\ee
Injecting into the exponential of the functional integral in Eq.~(\ref{corr1}) the relation~(\ref{exp}) and expanding the result at the linear order in $\delta H[\phi]$, the correlation function takes the form
\bea\label{gr1}
v(\br,\br')=\lan\phi(\br)\phi(\br')\ran_0-\left\{\lan\phi(\br)\phi(\br')\delta H[\phi]\ran_0-\lan\phi(\br)\phi(\br')\ran_0\lan\delta H[\phi]\ran_0\right\},\nonumber
\eea
with the perturbative part of the Hamiltonian functional 
\be
\delta H[\phi]=-\Lambda_s\int\frac{\mathrm{d}\br\mathrm{d}\bom}{4\pi}e^{-u\cos\theta}\left\{e^{iQ\left[\phi(\br)-\phi(\br+\ba)\right]}+\frac{Q^2}{2}\left[\phi(\br)-\phi(\br+\ba)\right]^2\right\}.\nonumber
\ee
In Eq.~(\ref{gr1}), the brackets denotes the average over potential fluctuations with respect to the quadratic Hamiltonian~(\ref{qu1}), that is 
\be\label{avwc}
\lan f[\phi]\ran_0\equiv\frac{1}{\sqrt{\rm{det}[v_0]}}\int \mathcal{D}\phi\;e^{-H_0[\phi]}f[\phi].
\ee
By evaluating the field-theoretic averages in Eq.~(\ref{gr1}) according to Eq.~(\ref{avwc}) and taking the Fourier transform of both sides of the equation, one gets after rather long algebra the Fourier-transformed propagator as
\be\label{gr2}
\tv(q_\perp,q_z)=\tv_0(q_\perp,q_z)+2Q^2\Lambda_s\int\frac{\mathrm{d}\bom}{4\pi}e^{-u\cos\theta}\left[1-e^{-Q^2\psi(\theta)}\right]\times\tv_0^2(q_\perp,q_z)\left\{1-\rm{J}_0(q_\perp a_\perp)\cos(q_za_z)\right\},
\ee
where we introduced the dipolar self-energy
\be\label{dipen}
\psi(\theta)=\int\frac{\mathrm{d}q_\perp q_\perp}{2\pi}\frac{\mathrm{d}q_z}{2\pi}\tv_0(q_\perp,q_z)\left\{1-\rm{J}_0(q_\perp a_\perp)\cos(q_za_z)\right\}
\ee
whose physical meaning will be emphasized below. Now, we note that the second term on the r.h.s. of Eq.~(\ref{gr2}) is the perturbation term that resulted from the cumulant expansion. Inverting this equation within the same approximation, one gets the Fourier-transformed inverse propagator as
\be\label{gr2II}
\tv^{-1}(q_\perp,q_z)=\tv_0^{-1}(q_\perp,q_z)-2Q^2\Lambda_s\int\frac{\mathrm{d}\bom}{4\pi}e^{-u\cos\theta}\left[1-e^{-\ell Q^2\psi(\theta)}\right]\left\{1-\rm{J}_0(q_\perp a_\perp)\cos(q_za_z)\right\}.
\ee

In the present work, we will not stay at the first order cumulant level where one should include hard-core interactions to regularize the solvent model but rather consider leading-order many-body effects beyond the mean-field (MF) limit. To this aim, we introduced in the argument of the exponential of Eq.~(\ref{gr2II}) the perturbative parameter $\ell$ that will allow to expand the Green's function perturbatively. The parameter $\ell$ will be set to unit at the end of the calculation. We also emphasize that since the perturbative parameter $\ell$ multiplies the dipolar self energy~(\ref{dipen}), its actual magnitude is determined by the strength of this self-energy. In other words, the perturbative expansion that will follow is expected to fail at strong external fields where the dipolar self-energy $\psi(\theta)$ becomes too large. In comparison with simulation data, the corresponding departure from the weak-coupling regime will be illustrated in Section~\ref{corr}.

 Before proceeding with the expansion,  we have to relate the dipolar fugacity to the solvent density.  The latter follows from the derivative of the grand potential $\Omega_G=-\ln Z_G$ with respect to the fugacity~\cite{Buyuk2014II},
\be\label{thermo1}
\rho_{s}=-\frac{\Lambda_s}{V}\frac{\partial\Omega_G}{\partial\Lambda_S}=\Lambda_S\int\frac{\mathrm{d}\bom}{4\pi}e^{-u\cos\theta}\lan e^{iQ\left[\phi(\br)-\phi(\br+\ba)\right]}\ran.
\ee 
Because we restrict ourselves to one-loop corrections linear in the electrostatic Green's function, evaluating the field theoretic average on the r.h.s. of Eq.~(\ref{thermo1}), we do not have to account for the contribution from the perturbative interaction term $\delta H[\phi]$ which brings second order cumulant corrections associated with direct particle-particle interactions. Computing this field-theoretic average with respect to the quadratic Hamiltonian $H_0[\phi]$ according to Eq.~(\ref{avwc}) and inverting the relation, one gets
\be\label{thermo2}
\Lambda_s=\rho_s/\int\frac{\mathrm{d}\bom}{4\pi}e^{-u\cos\theta} e^{-\ell Q^2\psi(\theta)},
\ee
where we introduced again the perturbative parameter $\ell$ that will allow to keep track of the perturbative order.  Inserting the fugacity in Eq.~(\ref{thermo2}) into Eqs.~(\ref{fr2}) and~(\ref{gr2II}), and Taylor-expanding Eqs.~(\ref{dipen}) and~(\ref{gr2II}) at the linear order in $\ell$, one obtains after some algebra the electrostatic kernel in the form
 \begin{widetext}
 \be\label{gr3}
 \tv^{-1}(q_\perp,q_z)=\tv^{-1}_{\rm{MF}}(q_\perp,q_z)+2Q^4\rho_s\left\{\lan\psi(\theta)\rm{J}_0(q_\perp a_\perp)\cos(q_za_z)\ran_\theta-\lan\psi(\theta)\ran_\theta\lan\rm{J}_0(q_\perp a_\perp)\cos(q_za_z)\ran_\theta\right\}.
 \ee
 \end{widetext}
 In Eq.~(\ref{gr3}), the dipolar self-energy reads
 \be\label{dipen2}
 \psi(\theta)=\int\frac{\mathrm{d}q_\perp q_\perp}{2\pi}\frac{\mathrm{d}q_z}{2\pi}\tv_{\rm{MF}}(q_\perp,q_z)\left\{1-\rm{J}_0(q_\perp a_\perp)\cos(q_za_z)\right\},
 \ee
 and the MF kernel has the form
 \be\label{grwc}
 \tv_{MF}^{-1}(q_\perp,q_z)=\frac{q^2}{4\pi\ell_B}+2Q^2\rho_sF(q_\perp,q_z).
 \ee
 Finally, the analytical form of the structure factor in Eq.~(\ref{grwc}) follows from the evaluation of the integral over dipolar rotations in Eq.~(\ref{f1}),
 \bea\label{f2}
 F(q_\perp,q_z)&=&1-\frac{f\cosh (g)\sin (f) +g\sinh (g)\cos (f)}{\left[f^2+g^2\right]\sinh (u)/u},\nonumber
 \eea
where we introduced the auxiliary functions
 \bea
 f&=&\frac{1}{\sqrt2}\left[\baq^2-u^2+\sqrt{(\baq^2-u^2)^2+4u^2\baq_z^2}\right]^{1/2}\\
 g&=&\frac{u\baq_z}{f}
 \eea
 whose dependence on the adimensional wavevectors $\baq_z=aq_z$ and $\baq_\perp=aq_\perp$ is implicit. 
  
Eqs.~(\ref{gr3})-(\ref{grwc}) are the main results of the present work. Before analyzing them, we note that the dipolar energy in Eq.~(\ref{dipen2}) exhibits an ultraviolet (UV) divergence due to the self-energy of the discrete charges on the dipole. This corresponds to the contribution from the first term inside the curly brackets of this equation. Although the divergence can be regularized by subtracting the self-energy of these charges in vacuum as in Ref.~\cite{Buyuk2014II}, this procedure is unnecessary in the present case. Indeed, the divergent contributions from the two terms in the  brackets of Eq.~(\ref{gr3})  cancel each other. Therefore, the term inside the brackets is finite.
 
\section{Dielectric saturation versus increment}
\label{res}

\subsection {Deriving dielectric permittivities and MF level dielectric anisotropy}

 In this part, we investigate the effect of the external field on the dielectric response properties of the present liquid model. To this aim, we consider the asymptotic large distance limit of the   
 kernel~(\ref{gr3}) where the latter takes the local form 
 \be\label{say}
 \tv^{-1}(q_\perp\to0,q_z\to0)=\frac{\e_\perp q_\perp^2+\e_z q_z^2}{4\pi\ell_B}.
 \ee
 In Eq.~(\ref{say}), the transverse and longitudinal components of the dielectric permittivity tensor read
 \bea\label{ep}
 \e_\perp&=&\e^{(0)}_{\perp}-\frac{(Q\kappa_sa)^2}{4}\left[\lan\psi(\theta)\sin^2\theta\ran_\theta-\lan\psi(\theta)\ran_\theta\lan\sin^2\theta\ran_\theta\right]\nonumber\\
 &&\\\label{ez}
 \e_z&=&\e^{(0)}_{z}-\frac{(Q\kappa_sa)^2}{2}\left[\lan\psi(\theta)\cos^2\theta\ran_\theta-\lan\psi(\theta)\ran_\theta\lan\cos^2\theta\ran_\theta\right],\nonumber\\ 
 &&
 \eea
 where we introduced the MF permittivities
 \bea\label{ep0}
 \e^{(0)}_{\perp}&=&1+\frac{(\kappa_sa)^2}{2}\frac{\rm{L}(u)}{u}\\
 \label{ez0}
 \e^{(0)}_{z}&=&1+\frac{(\kappa_sa)^2}{2}\left[1-\frac{2}{u}\;\rm{L}(u)\right],
 \eea
with the Langevin function $\rm{L}(u)=\coth(u)-1/u$ and the dipolar screening parameter $\kappa_s^2=8\pi Q^2\ell_B\rho_s$. First, one notices that Eq.~(\ref{ep0}) is similar to the Booth formula originally derived in Ref.~\cite{Booth1951}. Then, in the limit of vanishing external field, both permittivity components tend to the Debye-Langevin permittivity, i.e. $\e^{(0)}_{\perp,z}\to\e_w$ and $\e_{\perp,z}\to\e_w$ as $u\to0$, with
\be\label{dibulk}
\e_w=1+\frac{4\pi}{3}Q^2\ell_Ba^2\rho_s\approx77.
\ee

\begin{figure}
\includegraphics[width=.8\linewidth]{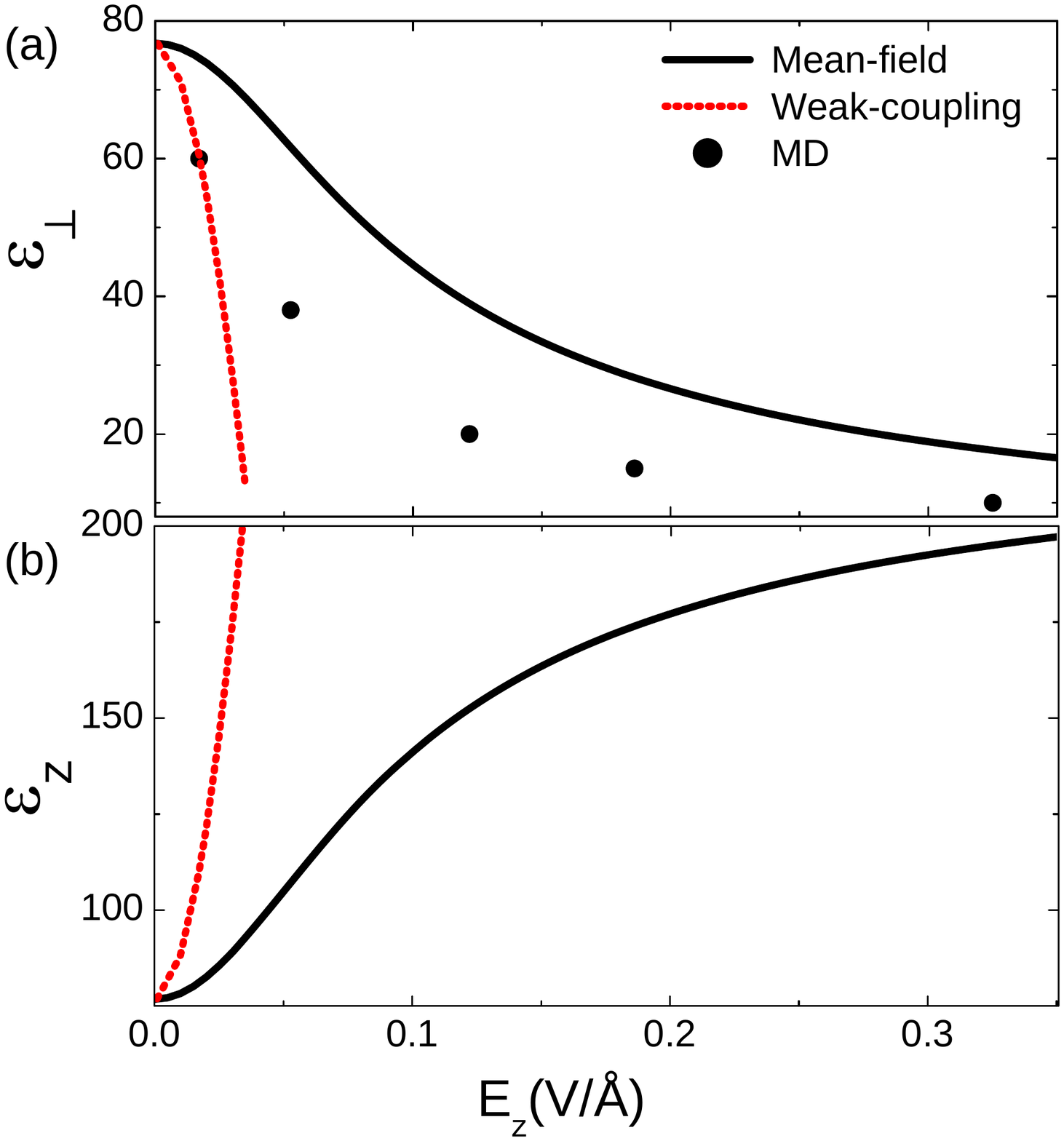}
\caption{(Color online) Components of the dielectric permittivity (a) perpendicular and (b) parallel to the external field as a function of the electric field $E_z$. Dashed red curves are the WC permittivities from  Eqs.~(\ref{ep}) and~(\ref{ez}), solid black curves denote the MF result of Eqs.~(\ref{ep0}) and~(\ref{ez0}), and dot symbols are MD data from Ref.~\cite{Yeh1999}. The model parameters are given in the caption of Fig.~\ref{Fig0}.}
\label{Fig1}
\end{figure}

We consider now the departure of the MF permittivity components~(\ref{ep0}) and~(\ref{ez0}) from the bulk value~(\ref{dibulk}) with the increase of the adimensional electric field $u$. First, we note that the presence of the two-component dielectric permittivity tensor in Eq.~(\ref{say}) indicates a dielectric anisotropy effect induced by the external field. As stated in the Introduction part, the breaking of the isotropic dielectric response indicates the inadequacy of considering external field effects on the dielectric response of polar liquids in terms of a single dielectric permittivity function~\cite{Kirkwood1939,Booth1951,Curry1992,Basu1994,Yeh1999,Gavryushov2007,Arzo2008,Szalai2009,Fulton2009}. Then, we display in Fig.~\ref{Fig1} the MF dielectric permittivities versus the external field. Increasing the field strength, the dielectric permittivity components $\e^{(0)}_\perp$ and $\e^{(0)}_z$ are shown to exhibit opposite behaviour. Namely, the dielectric permittivity $\e^{(0)}_\perp$ in the $x-y$ plane drops monotonically from $\e_w$ towards the vacuum permittivity, 
\be
\lim_{u\to\infty}\e^{(0)}_\perp=1,
\ee
which is the manifestation of the dielectric saturation effect perpendicular to the field. However, the permittivity $\e^{(0)}_z$ parallel to the field rises from $\e_w$ to  the limit 
\be
\lim_{u\to\infty}\e^{(0)}_z=1+4\pi Q^2\ell_Ba^2\rho_s>\e_w, 
\ee
which means that dielectric increment takes place along the field direction. We note that a similar dielectric anisotropy effect with the parallel component exceeding the bulk and perpendicular permittivities was observed in previous MD simulations of solvent molecules in nanoslits~\cite{hansim}. Again, the asymmetrical behaviour of the permittivity components displayed in Fig.~\ref{Fig1} shows the inconsistency of reducing external field effects on the dielectric response properties to a simple dielectric saturation effect without specifying the spatial direction with respect to the external field. The mechanism behind this dielectric anisotropy will be scrutinized with further detail in Section~\ref{corr}.

In order to map the results above onto the classical formulation of dielectric response in polar liquids, one can show that the MF permittivities~(\ref{ep0}) and~(\ref{ez0}) correspond exactly to the local permittivity tensor related to the variation of the average polarization field $\lan\bP\ran$,
\be\label{ten1}\label{ten1}
\frac{\e_{\alpha\beta}-1}{4\pi\ell_B}=\chi_{\alpha\beta}=\frac{\partial\lan P_\alpha\ran}{\partial E_\beta},
\ee
where the symbols $\alpha$ and $\beta$ denote the spatial directions $x,y,z$, and $\chi_{\alpha\beta}$ stands for the dielectric susceptibility tensor. Now, we note that the average polarization along the $\alpha$ direction follows from the functional derivative of the partition function~(\ref{zg1}) with respect to the external field,
\be
\lan P_\alpha\ran=\frac{1}{Z_G}\frac{\delta Z_G}{\delta E_\alpha(\br)}=-\frac{\delta\Omega_G}{\delta E_\alpha(\br)},
\ee
with the grand potential $\Omega_G=-\ln Z_G$. Hence, the susceptibility tensor in Eq.~(\ref{ten1}) reads
\be\label{ten2}
\chi_{\alpha\beta}=-\frac{\delta^2\Omega_G}{\delta E_\alpha(\br)\delta E_\beta(\br)}.
\ee
In the bulk liquid where the MF electrostatic potential is zero, the MF grand potential follows from Eq.~(\ref{HamFunc}) in the simple form $\Omega_G=-\Lambda_s\int\mathrm{d}\bom\;e^{-Q\bE\cdot\ba}/(4\pi)$. If one inserts this grand potential into Eq.~(\ref{ten2}) together with the relation $\Lambda_s=\rho_su/\sinh(u)$ that follows from the MF limit of Eq.~(\ref{thermo1}), the susceptibility tensor takes the form
\be\label{ten3}
\chi_{\alpha\beta}=Q^2\rho_s\frac{u}{\sinh(u)}\int\frac{\mathrm{d}\bom}{4\pi} e^{-Q\bE\cdot\ba}a_\alpha a_\beta.
\ee
By substituting into Eq.~(\ref{ten3}) the $x$ and $z$ components of the dipolar alignment vector, $a_x=a\sin\theta\cos\varphi$ and $a_z=a\cos\theta$ where $\varphi$ stands for the azimuthal angle in the $x-y$ plane, and carrying out the integrals over the dipolar rotation, one obtains the susceptibility components perpendicular and parallel to the external field as
\bea
\chi_{\perp}&=&\chi_{xx}=(Qa)^2\rho_s\frac{L(u)}{u}\\
\chi_{zz}&=&(Qa)^2\rho_s\left[1-\frac{2}{u}L(u)\right].
\eea
By using these susceptibilities in the first equality of Eq.~(\ref{ten1}), one gets exactly the MF permittivity components of Eq.~(\ref{ep0}) and~(\ref{ez0}).

\subsection{Dipolar correlation effects on dielectric anisotropy}
\label{corr}

We scrutinize in this part the role played by dipolar correlations on the dielectric anisotropy effect. To this aim, we reported in Fig.~\ref{Fig1}(a) numerical dielectric permittivity data obtained from MD simulations of Ref.~\cite{Yeh1999}. The comparison of the data with the MF curve shows that the MF theory underestimates the dielectric saturation effect~\cite{rem2}. In order to evaluate the relevance of solvent correlations, we also display in Figs.~\ref{Fig1}(a) and (b) the correlation-corrected permittivities of Eqs.~(\ref{ep})-(\ref{ez}). One notes that dipolar correlations strengthen the dielectric saturation perpendicular to the field and the dielectric increment parallel to the field, i.e. $\e_{\perp}<\e^{(0)}_\perp$ and $\e_{z}>\e^{(0)}_z$ for all finite values of the external field. Furthermore, correlation effects correct the MF result in the right direction for weak fields $E_z\ll0.1$ $\mbox{V}/\mbox{{\AA}}$ where the weak-coupling (WC) result $\e_\perp$ stays close to the simulation data.  However, the WC theory fails at external fields beyond this regime and overestimates correlations, a limitation known from solvent-implicit one-loop theories of charged liquids~\cite{Podgornik1988,Netz2000,Lau2008,Buyuk2012,Buyuk2013}. The failure of the present one-loop theory corresponds to the departure from the WC regime where the characteristic energy associated with dipolar rotations under the external field $E_z$  becomes much larger than the thermal energy $k_BT$. Consequently, the expansion of the grand potential in terms of this characteristic energy becomes invalid in this regime.  \textcolor{black}{Finally, for the sake of a qualitative mapping to the Madden-Kivelson theory of dielectric response, it is noteworthy to mention that many-body interactions can be alternatively taken into account in terms of the Kirkwood factor~\cite{Madden,Netz2014}. The latter is defined as the ratio of the correlation corrected and single-particle (or mean-field) susceptibilities, that is $g=\chi/\chi_{MF}$. The above discussion of Fig.~\ref{Fig1} shows that at finite external fields, this factor is lower than one for the perpendicular component (i.e. $g_\perp<1$) and larger than unit for the parallel component  (or $g_{zz}>1$).}
\begin{figure}
\includegraphics[width=.6\linewidth]{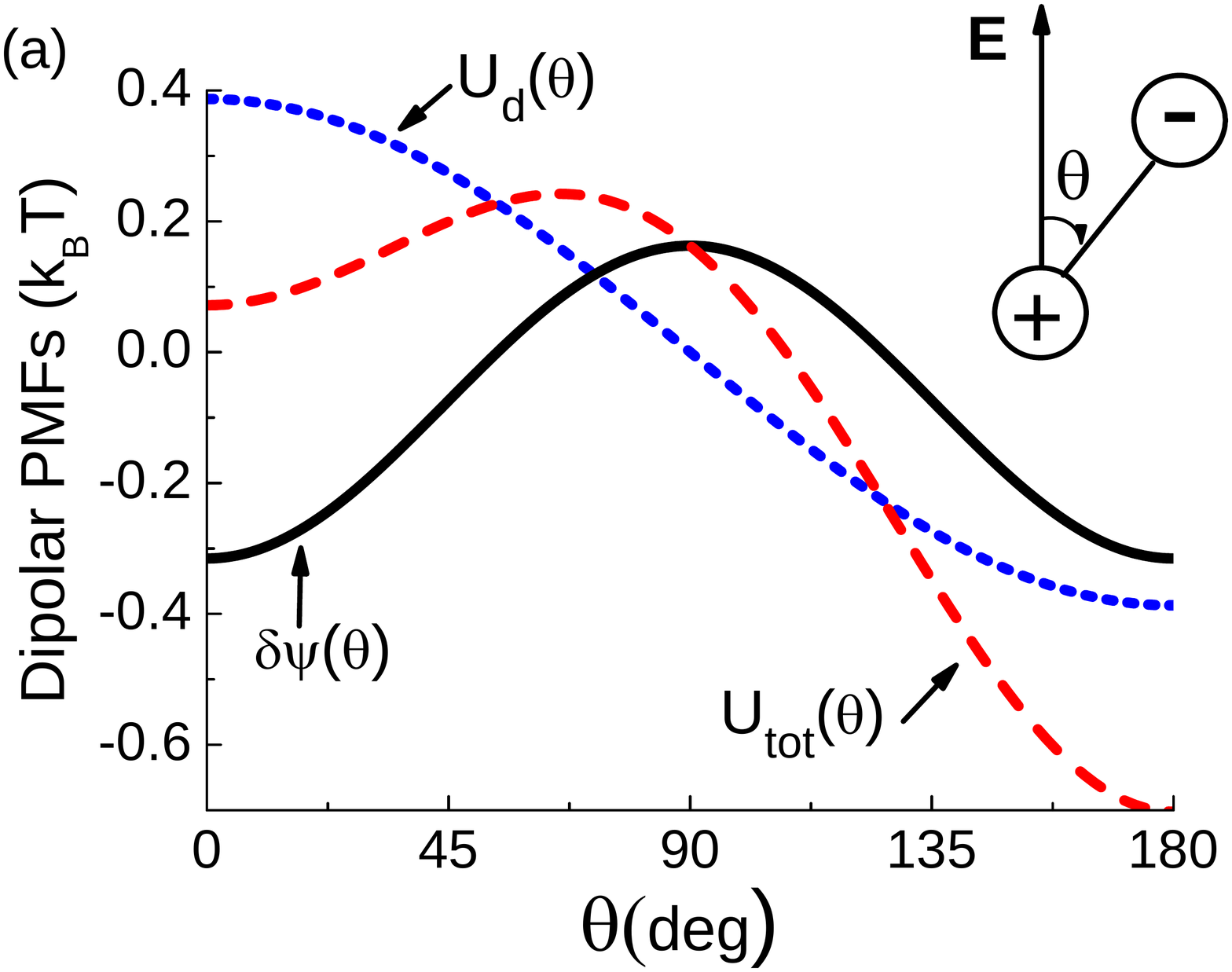}
\includegraphics[width=.6\linewidth]{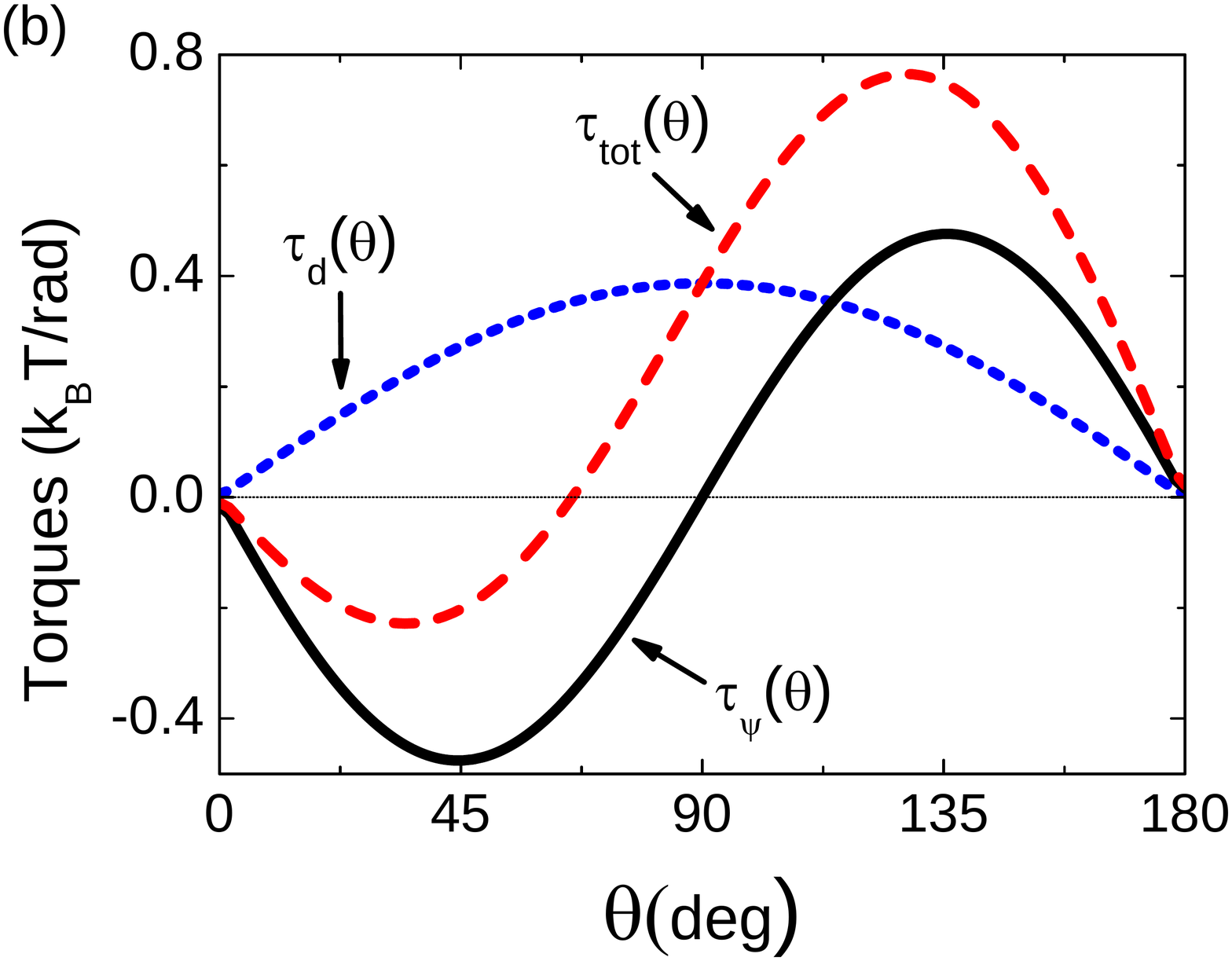}
\caption{(Color online) (a) Dipole-field coupling energy $U_d(\theta)=-QaE_z\cos\theta$, reduced dipolar self-energy $\delta\psi(\theta)=\psi(\theta)-\lim_{E_z\to0}\psi(\theta)$, and total dipolar energy $U_{\rm{tot}}(\theta)=U_d(\theta)+\delta\psi(\theta)$ as a function of the angle $\theta$ between the dipole and the external field. (b) Torque applied  to the dipole by the external field $\tau_d(\theta)=QaE_z\sin\theta$, electrostatic many-body forces $\tau_\psi(\theta)=-\psi'(\theta)$,  and total force $\tau_{\rm{tot}}(\theta)=-U_{\rm{tot}}'(\theta)$. The magnitude of the external field is $E_z=0.01$ $\mbox{V}/\mbox{{\AA}}$ in both figures. The other model parameters are given in the caption of Fig.~\ref{Fig0}.}
\label{Fig2}
\end{figure}

In order to scrutinize in further detail the mechanisms driving the dielectric anisotropy and the role played by correlations, we plotted in Fig.~\ref{Fig2}(a) the electrostatic potentials experienced by dipoles. First,  the dipole-field coupling potential $U_{\rm{d}}(\theta)$ already present at the MF level is seen to decrease from $\theta=0{\degree}$ to $\theta=180{\degree}$, favouring the polarization of dipoles along the field with the positive charges pointing the field direction. This effect is responsible for the saturation of the MF dielectric permittivity component perpendicular to the field and the increment of the component parallel to the field (see Fig.~\ref{Fig1}). We consider now the dipolar self energy~(\ref{dipen2}) accounting for many-body effects rescaled by its value at vanishing field,  
\be\label{disel}
\delta\psi(\theta)=\psi(\theta)-\lim_{E_z\to0}\psi(\theta),
\ee
where the subtracted term has no $\theta$-dependence~\cite{rem1}. The potential $\delta\psi(\theta)$ accounts for the energetic cost associated with the perturbation of the solvent cloud surrounding a dipole rotated by $\theta$. Fig.~\ref{Fig2}(a) shows that this potential rises from $\theta=0{\degree}$ to $\theta=90\degree$ where it reaches a peak and drops symmetrically towards $\theta=180\degree$. First, this indicates that the electrostatic force on the dipole induced by correlation effects does not distinguish between the parallel and antiparallel alignments with respect to the external field. Then, correlations lead to an energetic penalty  for the alignment of the dipole perpendicular to the field. As a result, in Fig.~\ref{Fig2}(a), dipolar correlations are shown to lower the minimum of the total potential of mean force $U_{\rm{tot}}(\theta)=U_{\rm{d}}(\theta)+\delta\psi(\theta)$ at $\theta=180\degree$, strengthening the tendency of the dipolar alignment along the field direction. This explains the intensification of the dielectric decrement and increment effects by dipole correlations in Figs.~\ref{Fig1}(a) and (b), respectively. Moreover, Fig.~\ref{Fig2}(a) shows that the configuration with the dipolar angle $\theta=0\degree$ that is an unstable point at the MF level becomes metastable in the presence of dipolar correlations, with the apparition of a new unstable point slightly below $\theta=90\degree$.  This prediction can be tested in MD simulations where the metastable equilibrium point at $\theta=0\degree$ is expected to trap the solvent molecules for a while.  

As an alternative to the potential description above, one can analyze the effect of correlations on the dipolar orientation in terms of the forces acting on the solvent molecules. To this aim, we plotted in Fig.~\ref{Fig2}(b) the electrostatic torques on dipoles associated with the potentials of Fig.~\ref{Fig2}(a). First, the torque resulting from the MF-level dipole-field coupling has the well-known form $\tau_d(\theta)=-U'_d(\theta)=QaE_z\sin\theta$ (dotted blue curve). This positive torque with the maximum magnitude at the angle $\theta=90\degree$ forces dipoles to align with the electric field, resulting in the dielectric anisotropy effect that we have scrutinized in detail. Then, the torque $\tau_\psi(\theta)=-\psi'(\theta)$ associated with correlations is positive for $\theta<90\degree$ and negative for $\theta>90\degree$ (see the solid black curve). Thus, in agreement with Fig.~\ref{Fig2}(a), many body forces tend to drive dipoles towards the equilibrium points at $\theta=0\degree$ and $\theta=180\degree$. Furthermore, the correlation-induced torque on the dipoles reaches its largest magnitude at the angles $\theta=45\degree$ and $\theta=135\degree$. As a result, the total torque $\tau_{\rm{tot}}(\theta)=-U_{\rm{tot}}'(\theta)$ on the solvent molecules has a considerably larger magnitude than the MF one beyond the unstable equilibrium point $\theta\simeq68\degree$, indicating the amplification of the dipolar alignment tendency along the field direction by dipole-dipole correlations.

\section{Summary and Conclusions}

In this work, we characterized external field effects on the dielectric response of a polar solvent. Within a grand-canonical calculation, we developed the first electrostatic theory of dielectric anisotropy induced by externally applied electric fields. The calculation of the two-point correlation function provided us with the electrostatic kernel of the liquid, which allowed an unambiguous derivation of the liquid dielectric permittivity. This point is the main advantage of the present field-theoretic approach over previous theories~\cite{Kirkwood1939,Booth1951,Curry1992,Basu1994,Yeh1999,Gavryushov2007,Arzo2008,Szalai2009,Fulton2009}. Indeed, the macroscopic equations such as Eq.~(\ref{bo}) used by previous works to define the solvent dielectric permittivity prevented them from identifying the dielectric anisotropy effect. The consideration of solvent correlation effects neglected by early works is an additional improvement brought by our formalism.

We found that that the external field breaks the dielectric isotropy of the solvent, which translates into a tensor form of the dielectric permittivity. At the MF level, due to the dipole-field coupling polarizing solvent molecules along the field (i.e. the configuration with $\theta=180\degree$ in Fig.~\ref{Fig2}(a)), the component of the permittivity tensor perpendicular to the field exhibits a dielectric saturation effect while dielectric increment takes place parallel to the field.  This dielectric anisotropy shows that the classical approach that consists in considering external field effects in terms of a scalar dielectric permittivity is not consistent~\cite{Kirkwood1939,Booth1951,Curry1992,Basu1994,Yeh1999,Gavryushov2007,Arzo2008,Szalai2009,Fulton2009}.  We considered as well the role played by solvent correlations on the dielectric anisotropy.  Because the energetic cost associated with the perturbation of the solvent cloud surrounding a central dipole is the highest when the latter is perpendicular to the electric field, many-body interactions favour dipolar alignment along the field and intensify the MF-level dielectric anisotropy. Furthermore, by comparison with MD simulation data of Ref.~\cite{Yeh1999}, MF dielectric response was shown to underestimate the dielectric saturation. The inclusion of dipole correlations corrects this error at weak fields but the present weak-coupling theory overestimates correlation effects beyond this regime. We also found that due to solvent correlations, the antiparallel dipolar alignment $\theta=0\degree$ that is unstable within the MF theory becomes a metastable configuration. This point can be verified in future MD simulations since the metastable state is expected to trap solvent molecules temporarily. 

We emphasize that the present solvent model lacks several features of biological solvents that should be considered in future works. First,  the model neglects hard-core interactions between solvent molecules. Considering the high density of water $\rho_{sb}=55$ M, this complication can quantitatively change the results of our article. In principle, core interactions between solvent molecules can be incorporated in terms of repulsive Yukawa interactions as in Ref.~\cite{jstat}. However, in order to cover finite size effects in a bulk liquid, one should expand the grand potential up to the second cumulant order. This step will bring numerical complexity that would shadow the analytical simplicity of the present model.  Then, we neglected as well multipolar moments of water molecules that can be added in an improved model in the form of linear multipoles~\cite{NLPB1}. We also omitted herein the hydrogen bonding between solvent molecules. Since hydrogen bonds strengthen the cooperativity in the liquid, their inclusion is expected to intensify the dielectric anisotropy effect. A systematic comparison with MD simulation will be necessary in order to evaluate the importance of these features on the key effect of dielectric anisotropy revealed by our formalism. Finally, we note that the present work focused exclusively on the local limit of the dielectric permittivity function. The non-locality of the solvent dielectric response embedded in the kernel equation~(\ref{gr3})  should be investigated in a future work.


\begin{thebibliography}{99}
\bibitem {rem1} The limit $\lim_{u\to0}\psi(\theta)$ corresponds to the electrostatic hydration energy of a finite size dipole in an isotropic environment where there is no privileged direction.
\end{thebibliography}
\end{document}